\documentclass[epj]{svjour}
\usepackage{graphics}
\usepackage{psfig}

\begin{document}
%\psdraft
\title{
Spin susceptibility in small Fermi energy systems:
effects of nonmagnetic impurities} 
\subtitle{}
\author{E.  Cappelluti\inst{1}
\and C. Grimaldi\inst{2} \and L. Pietronero\inst{1,3}}
\institute{Dipartimento di Fisica, Universit\'{a} ``La Sapienza", 
P.le A.  Moro 2, 00185 Roma, and INFM Roma1, Italy 
\and Ecole Polytechnique F\'ed\'erale de Lausanne, IPR-LPM,
CH-1015 Lausanne, Switzerland
\and Istituto di Acustica ``O.M. Corbino'',
CNR, Area di Ricerca Tor Vergata, Roma, Italy}
\date{\today/ \mbox{}}
\abstract{
In small Fermi energy metals, disorder
can deeply modify superconducting state properties 
leading to a strong suppression of the critical temperature $T_c$.
In this paper, we show that
also normal state properties can be seriously
influenced by disorder when the Fermi energy $E_{\rm F}$ is
sufficiently small. We calculate the normal state spin
susceptibility $\chi$ for a 
narrow band electron-phonon coupled metal as a function of
the non-magnetic impurity scattering rate $\gamma_{\rm imp}$.
We find that as soon as
$\gamma_{\rm imp}$ is comparable to $E_{\rm F}$, $\chi$
is strongly reduced with respect to its
value in the clean limit. The effects of the electron-phonon interaction
including the
nonadiabatic corrections are discussed. Our results strongly
suggest that the recent finding 
on irradiated MgB$_2$ samples can be naturally
explained in terms of small $E_{\rm F}$ values
associated with the $\sigma$-bands of the boron plane,
sustaining therefore the hypothesis 
that MgB$_2$ is a nonadiabatic metal. 
}
\maketitle

\section{Introduction}

Scattering from weak disorder or diluted non magnetic impurities
plays a marginal
role on many thermodynamics quantities of conventional metals.
Most peculiar is the absence of any reduction on the critical
temperature $T_c$ in conventional isotropic $s$-wave superconductors
as stated by the Anderson's theorem  and as confirmed
by several experimental measurements \cite{anderson}.

This insensitivity stands out in particular in comparison with
$d$-wave superconductors where the strong an\-iso\-trop\-y
of the order parameter
leads to a suppression of $T_c$ \cite{dwave}.
In that case for instance the reduction on
$T_c$ upon disorder can give
qualitative information of the microscopic characteristic on the pairing
($d$- vs. $s$- wave symmetry, local vs. long-ranged interaction, etc \ldots)
\cite{flatte}.

A conventional role of nonmagnetic impurities is recently questioned
in some high-$T_c$ superconductors, as MgB$_2$ and fullerene compounds.
In these materials a notable reduction of $T_c$ upon disorder
has been reported in spite of the $s$-wave symmetry of
both of them \cite{watson,karkin}.
Quite remarkable is also the reduction of the density of states (DOS)
inferred by NMR measurements of
the nuclear spin-lattice relaxation rate $T_1$
[$1/T_1 T \propto N_0^2$ where $N_0$ is the electronic density of
states at the Fermi level 
and $T$ is the temperature].
In Ref. \cite{gerashenko}
a reduction of 62 \% of $1/T_1 T$ upon disorder
was reported by $^{11}$B NMR measurements
in contrast with the conventional theory of non magnetic impurity
scattering which would predict no effect of the magnetic susceptibility.
An additional puzzling feature is the discrepancy between
spin-lattice relaxation rate measurements performed
on $^{11}$B NMR and on $^{25}$Mg. No reduction of $1/T_1 T$ was
indeed observed on magnesium atoms. The authors of
Ref. \cite{gerashenko} speculate this difference could be related to the
different nature of the electronic states: magnesium atoms would mainly
probe the $\pi$ bands of MgB$_2$ through the hybridization of B($2p_z$)
orbitals with Mg(s) states, while the spin-lattice relaxation rate
on the boron is expected to be very sensitive to the twodimensional
$\sigma$ bands formed by B($2p_x 2p_y$).

The evidences of anomalous effects of disorder and non magnetic
impurities in these systems prompt thus some intriguing open questions:
$i$) which is the origin of the suppression of $T_c$ in $s$-wave
systems as MgB$_2$ and fullerenes? $ii$) which is the origin of the
reduction of the density of states as probed by
spin-lattice relaxation rate $1/T_1 T$ measurements?
$iii$) which is the origin of the different behaviour of Mg and B
NMR measurements? 

The point ($i$) was previously addressed in Ref. \cite{sgp}
in the context of a nonadiabatic theory of superconductivity \cite{gps} where
impurity effects in the nonadiabatic channels were shown
to suppress $T_c$ even for purely isotropic $s$-wave
superconductors.
In this paper we extent our analysis to the spin susceptibility.
In particular
we show that a possible unifying explanation of all this
complex anomalous scenario could come from taking into account
in a coherent way the small Fermi energy nature of these materials.
This is clearly unavoidable in C$_{60}$ compounds where the narrow
bandwidth of the $t_{1u}$ bands (but also the $h_{u}$ bands for hole doped C$_{60}$)
results in a Fermi energy $E_{\rm F} \sim 0.25$ eV \cite{gunny}. This is also true
in MgB$_2$ where the low hole filling of the 2D $\sigma$ bands leads
to $E_{\rm F}^\sigma \sim 0.4-0.6$ eV \cite{an,kortus}. These values of $E_{\rm F}$
are at least one order of magnitude less than in common metals and in conventional
superconductors. The discrepancy between Mg and B NMR measurements can
be thus related to the probing of different bands ($\pi$ on Mg, 
$\sigma$ on B), and, in the last analysis, to the different magnitude
of the Fermi energies ($E_{\rm F}^\pi \sim 5$ eV $\gg E_{\rm F}^\sigma$).

On microscopic grounds, small Fermi energy effects are operative as soon
as $E_{\rm F}$ becomes of the same order of the other relevant energy
scales. For an electron-phonon system in the presence of
non magnetic impurities as we consider here, $E_{\rm F}$ should be thus
compared with the characteristic phonon energy scale $\omega_{\rm ph}$
and with the impurity scattering rate $\gamma_{\rm imp}$.
The breakdown of the adiabatic hypothesis
($E_{\rm F} \gg \omega_{\rm ph}$) in a small Fermi energy system implies
the onset of new channels of electron-phonon interaction which
need to be taken into account. On the other hand the finiteness of the
ratio $\gamma_{\rm imp}/E_{\rm F}$ gives rise to anomalous impurity
effects which have to be analyzed in the presence of the same
electron-phonon interaction since
electron, phonon and impurity energy scales could be all of
the same magnitude: $E_{\rm F} \sim \omega_{\rm ph}
\sim \gamma_{\rm imp}$.

\section{The model}

In this section, we derive the electron spin susceptibility
by employing the Baym-Kadanoff technique which permits to derive,
within a conserving theory,
higher order response functions as functional derivatives
of the single particle Green's function
in the presence of an external field \cite{baym}.
This approach is thus an appropriate starting point
to study small Fermi energy systems
where the violation of the Migdal's theorem valid
for $E_{\rm ph} \gg \omega_{\rm ph}$ requires a generalization
of the conventional theory in the nonadiabatic regime.

Objects of our investigation is the non magnetic impurity effects
on the spin susceptibility in small Fermi energy systems
in the presence of a sizable electron-phonon interaction. 
NMR techniques can probe
the electron density of states by means of different ways.
Most direct is the evaluation of the static uniform
limit $\chi$ of the generalized electron
spin susceptibility $\chi({\bf q},\omega)$:
\begin{equation}
\chi = \lim_{{\bf q} \rightarrow 0} \lim_{\omega \rightarrow 0}
\chi({\bf q},\omega)
\label{pauli}
\end{equation}
which, for a non interacting system with large Fermi energy,
is simply $\chi \propto N_0$.
Electron-electron exchange interaction gives rise however
to the so called Stoner enhancement: $\chi \propto N_0/(1-I)$
($I$ being the Stoner factor).
Experimentally the static uniform limit $\chi$ of
electron spin susceptibility can be measured
by a proper analysis of the Knight shift after the orbital
contribution is subtracted.

Similar information are obtained by 
spin-lattice relaxation rate $T_1$ which can be also mainly related,
after subtraction of orbital terms,
to the electron spin susceptibility through the relation:
\begin{equation}
\frac{1}{T_1T} \propto
\lim_{\omega \rightarrow 0}
\sum_{\bf q} A^2({\bf q}) \frac{\chi({\bf q},\omega)}{\omega},
\label{t1}
\end{equation}
where $A({\bf q})$ is the form factor relative
to the particular nucleus. As pointed out in the introduction, 
$1/T_1 T \propto N_0^2$ in large Fermi energy systems.

In the following we focus on the static uniform spin susceptibility
$\chi$
which permits a more direct comparison with the density of states
and which is only slightly affected by different form factors $A({\bf q})$.
As it will be clear in the following the anomalous effects of non magnetic
impurities are essentially related to the similar energy scales
of $\gamma_{\rm imp}$, $E_{\rm F}$ and $\omega_{\rm ph}$. In this situation
impurity scattering leads to an effective renormalization of the
generalized spin susceptibility
which is expected to appear in similar way
both in the static uniform limit $\chi$
and in the spin-lattice relaxation rate.
In this perspective the reduction of the magnetic susceptibility
upon disorder pointed out by NMR technique should be read more
as an anomalous renormalization effect appearing in small
Fermi energy system than as a real reduction of the density of states.

In Quantum Field Theory
the electron spin susceptibility is usually related
to the one particle Green's function $G$ through the
relation:\cite{note-localeffects}
\begin{equation}
\chi(T)=
-2\mu_{\rm B}^2T\sum_n\sum_{{\bf k}}
G({\bf k},n)^2\Gamma({\bf k},n),
\label{chi}
\end{equation}
where $G({\bf k},n)$ is the electron propagator at finite temperature
expressed in
Matsubara frequencies
\begin{equation}
G^{-1}({\bf k},n)=i\omega_n-\epsilon({\bf k})+\mu-
\Sigma({\bf k},n),
\label{green}
\end{equation}
and $\Gamma({\bf k},n)$ is the spin vertex function.
The Baym-Kadanoff formalism provides a powerful technique to related
the spin vertex function $\Gamma({\bf k},n)$ to $G({\bf k},n)$.
Following the standard procedure we generalize
the Green's function in Eq. (\ref{green}) in the presence
of an external magnetic field $h$:
\begin{equation}
G^{-1}_\sigma({\bf k},n)=i\omega_n-\epsilon({\bf k})+\mu
+h\sigma-\Sigma_\sigma({\bf k},n).
\label{greenh}
\end{equation}
The spin vertex function is thus obtained
as functional derivative of the Green's function $G_\sigma$
in the presence of the external magnetic field for $h \rightarrow 0$ \cite{cgp}:
\begin{eqnarray}
\Gamma({\bf k},n)&=&\frac{1}{2}\sum_{\sigma}\sigma
\left[\frac{d G_{\sigma}^{-1}({\bf k},n)}{dh}\right]_{h=0} 
\nonumber\\
&=&
1-\frac{1}{2}\sum_{\sigma}\sigma
\left[\frac{d \Sigma_{\sigma}({\bf k},n)}{dh}\right]_{h=0}.
\label{gammaspin}
\end{eqnarray}
The set of Eqs. (\ref{chi})-(\ref{gammaspin}) defines a self-consistent
method to obtain the spin susceptibility from the knowledge of
the self-energy. The complex nature of the interactions in the systems
is thus hidden in the specific form of the self-energy which needs
to be explicitly provided.

\begin{figure}
\centerline{\psfig{figure=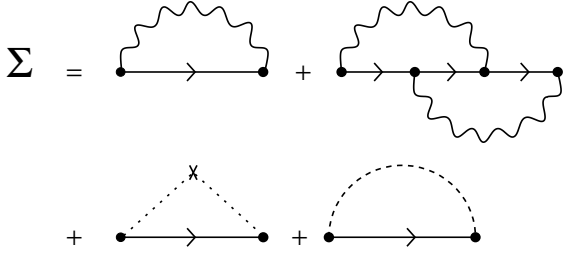,width=8cm}}
\vspace{1mm}
\caption{Diagrammatic picture of the electron self-energy.
Legend of the pictorial elements:
electrons (solid line), phonons (wavy lines), elastic scattering
(dotted lines) with dilute non magnetic impurities (cross),
electron-electron repulsion (dashed line).}
\label{f-self}
\end{figure}

In order to investigate the interplay between non magnetic impurities and
the electron-phonon interaction
in small Fermi energy systems an appropriate approach is
the nonadiabatic theory which accounts for the
additional interaction channels arising 
when $E_{\rm F} \sim \omega_{\rm ph}$.
The formal derivation of the nonadiabatic theory has been already
presented in several papers where we refer for more
details \cite{sgp,gps,cgp}.
Here we focus on the role of non magnetic impurities.
In the spirit of the Baym-Kadanoff theory our starting point
will be the self-energy which is diagrammatically depicted in
Fig. \ref{f-self}.
The first two diagrams represent the electron-phonon interaction
in nonadiabatic regime ($E_{\rm F} \sim \omega_{\rm ph}$) including
the first order vertex processes; the third diagram is the 
self-energy in Born approximation for impurities of density $n_{\rm imp}$
interacting with electrons via a scattering potential $v_{\rm imp}$.
The last diagram is the exchange electron-electron
interaction: this term is just a constant and does not play any role
in the self-energy, but it gives rise to the Stoner factor
in the spin susceptibility.

Fig. \ref{f-self} defines in an unambiguous way the self-energy
and the one particle properties of the system. Standard procedure
in isotropic materials is to replace the self-en\-er\-gy $\Sigma({\bf k},n)$
with its Fermi surface average:
$\Sigma({\bf k},n) \rightarrow \Sigma(n) \equiv 
\langle \langle \Sigma({\bf k},n) \rangle \rangle_{\rm FS}$.
It is convenient to take into account self-energy effects
in the electronic Green's function is to introduce the
renormalized Matsubara frequencies defined as $iW_n = i\omega_n -\Sigma(n)$.
In addition, for sake of simplicity we consider a half-filled band
with bandwidth $E$ and constant density of states
$N(\epsilon) = N_0$ [$-E/2 \le \epsilon \le E/2$]. The parameter
$E/2$ represents thus the Fermi energy $E_{\rm F}=E/2$.
Within these assumptions the analytic expression of the renormalized
Matsubara frequencies $W_n$ involving the self-energy
depicted in Fig. \ref{f-self} reads:
\begin{eqnarray}
W_n &=& \omega_n
-2T \sum_M V(n,m) \arctan\left(\frac{E_{\rm F}}{W_m}\right)
\nonumber\\
&& +2\gamma_{\rm imp} 
\arctan\left(\frac{E_{\rm F}}{W_n}\right),
\label{selfeq}
\end{eqnarray}
where
\begin{equation}
\label{V1}
V(n,m)=\lambda D(n-m)[1+\lambda P(Qc;n,m)]
\end{equation}
is the nonadiabatic electron-phonon kernel
appearing in the self-energy equation and where
we have neglected the
electron-electron exchange interaction which leads just to a constant
term. In Eq. (\ref{V1}), $D(n-m)$ is the phonon propagator which
for a single Einstein mode $\omega_{\rm ph} = \omega_0$ reduces
simply to $D(n-m)=-\omega_0^2/[(\omega_n-\omega_m)^2+\omega_0^2]$ and
$P(Q_c;n,m)$ is the vertex function \cite{sgp,cgp}:
\begin{eqnarray}
\label{vertex2}
&P&(Q_c;n,m)   =   -T\sum_lD(n-l)
\left\{\frac{}{}B(n,m,l)\right. \nonumber \\
& + & \frac{A(n,m,l)-B(n,m,l)(W_l-W_{l-n+m})^2}
{(2E_FQ_c^2)^2} \nonumber \\
& \times & \left.\left[R(Qc;n,m,l)-1-\log\left(\frac{1+R(Q_c;n,m,l)}{2}\right)
\right]\right\} , \nonumber \\
\end{eqnarray}
where 
\begin{eqnarray}
\label{a}
A(n,m,l) & = & (W_l-W_{l-n+m})\left[\arctan\left(\frac{E_F}{W_l}\right)
\right. \nonumber \\
& - &\left.\arctan\left(\frac{E_F}{W_{l-n+m}}\right)\right] ,
\end{eqnarray}
\begin{eqnarray}
\label{b}
B(n,m,l) & = & (W_l-W_{l-n+m})\frac{E_FW_{l-n+m}}
{[E_F^2+W_{l-n+m}^2]^2} \nonumber \\
& - & \frac{E_F}{E_F^2+W_{l-n+m}^2} ,
\end{eqnarray}
\begin{equation}
\label{r}
R(Q_c;n,m,l)=\sqrt{1+\left(\frac{4E_FQ_c^2}{W_l-W_{l-n+m}}\right)^2} .
\end{equation}
The dimensionless parameter $Q_c=q_c/2k_{\rm F}$, where $k_{\rm F}$ is the Fermi momentum,
takes into account the upper cutoff $q_c$ for the momentum transfer in the
electron-phonon interaction. This cutoff has been introduced to simulate a
momentum dependent renormalization due to possible
strong electronic correlations \cite{kulic}.
For weak correlated metals $Q_c\simeq 1$, while $Q_c\ll 1$ when correlation is strong.
As we are going to see, the parameter $Q_c$ plays only a marginal
on the static spin susceptibility, whereas it strongly affects
the superconducting critical temperature \cite{gps}.

In the formula for $W_n$, Eq. (\ref{selfeq}), $\gamma_{\rm imp}$ is the
impurity scattering rate which in the Born approximation reduces to
$\gamma_{\rm imp}=\pi n_{\rm imp}N_0 v_{\rm imp}^2$ \cite{rick}. This expression holds true for
low values $n_{\rm imp}$ of impurity concentrations and
weak scattering potential $v_{\rm imp}$. An expression of 
$\gamma_{\rm imp}$ valid also for strong, but diluted, impurity interactions is
provided by the $T$-matrix approximation: 
$\gamma_{\rm imp}=
\pi n_{\rm imp}N_0 v_{\rm imp}^2/[1+(\pi N_0 v_{\rm imp})^2]$.

\begin{figure}
\centerline{\psfig{figure=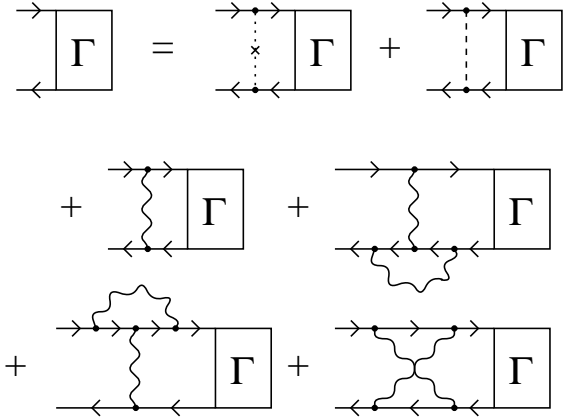,width=8cm}}
\vspace{1mm}
\caption{Diagrammatic picture of spin vertex function $\Gamma$.
See caption of \protect Fig. \ref{f-self} for a legend
of the pictorial elements.}
\label{f-gamma}
\end{figure}

Using the Baym-Kadanoff formalism we are able
to obtain also an analytic expression
for the spin vertex function $\Gamma({\bf k},n)$.
The diagrammatic expression of $\Gamma({\bf k},n)$ corresponding
to the self-energy depicted in Fig. \ref{f-self} is shown
in Fig. \ref{f-gamma}.
In the isotropic case we have considered here
we can replace also the spin vertex function $\Gamma({\bf k},n)$
with its Fermi surface average $\Gamma(n)$.
The momentum average of the spin vertex implies that the
momentum correlations between spin up electrons and spin down holes
are taken into account only at a mean level through the parameter $Q_c$.
This will be a poor approximation when the dispersion
of collective modes is investigated, while it is
expected to not affect in a qualitative way
the static uniform spin susceptibility.
Disregarding the momentum dependence of
$\Gamma({\bf k},n)$ and we get thus:
\begin{eqnarray}
\Gamma (n)&=&1 + T\sum_{m}\left[I+V_\Gamma(n,m)\right]
\frac{2E_F}{W_m^2+E_F^2}\Gamma(m)
\nonumber\\
&&
-\gamma_{\rm imp}\frac{2E_F}{W_n^2+E_F^2}\Gamma(n),
\label{gammavertex}
\end{eqnarray}
where the quantity $I=N_0 U$ is the Stoner factor arising from the
electron-electron exchange interaction and the last term
comes from the impurity scattering processes. 
Moreover
\begin{eqnarray}
\label{ave6}
V_\Gamma(n,m)&=&\lambda D(n-m)\left[
1+2 \lambda P(Q_c;n,m)\right]  \nonumber \\
&+&\lambda^2 C(Q_c;n,m)
\end{eqnarray}
describes
the electron-phonon processes in nonadiabatic regime which include
electron-pho\-non vertex $P(Q_c;n,m)$ given by Eqs.(\ref{vertex2}-\ref{r}) the 
and cross diagrams:
\begin{eqnarray}
\label{crossb}
&C&\!(Q_c;n,m) = T\!\sum_l D(n-l)D(l-m) \nonumber \\
& \times &\!\left\{2B(n,m,l)+\arctan\left(\frac{4E_FQ_c^2}
{|W_l-W_{n+m-l}|}\right)\right. \nonumber \\
& \times &\left.
\frac{A(n,m,l)-B(n,m,l)(W_l-W_{n+m-l})^2}
{2E_FQ_c^2|W_l-W_{n+m-l}|}\right\} ,
\end{eqnarray}
where $A(n,m,l)$ and $B(n,m,l)$ are given by equations
(\ref{a}) and (\ref{b}), respectively.
Note that 
$V_\Gamma(n,m)$ is deep\-ly different from $V(n,m)$
since the first describes electron-phonon scattering in the spin
electron-hole channel, and the second one the electron-phonon interaction
in the single particle propagator.

\section{Results and discussion}

Eqs. (\ref{selfeq})-(\ref{crossb})
can be solved in a self-consistent iterative way to obtain
$W_n$ and $\Gamma(n)$.
Eq. (\ref{chi}), in its isotropic form:
\begin{equation}
\chi(T)=
\chi_{\rm P} T\sum_n
\frac{2E_{\rm F}}{W_n^2+E_{\rm F}^2}
\Gamma(n),
\label{chi-iso}
\end{equation}
provides finally the spin susceptibility as function
of ge\-ner\-ic impurity scattering rate $\gamma_{\rm imp}$,
electron-phonon coupling constant $\lambda$, adiabatic ratio
$\omega_0/E_{\rm F}$ and momentum cut-off $Q_c$.
Here $\chi_{\rm P}$ is the free electron Pauli spin susceptibility
$\chi_{\rm P} = 2\mu_{\rm B}^2 N_0$.

In order to point out the role of small Fermi energy in
the impurity scattering effects on the spin susceptibility,
we consider for the moment the simple case of no electron-phonon
interaction ($\lambda=0$). In this case the only energy scales
in the system are $\gamma_{\rm imp}$ and $E_{\rm F}$.
From Eq. (\ref{chi-iso}), equation (\ref{gammavertex})
has thus the simple self-consistent
solution as function of the spin susceptibility itself:
\begin{equation}
\Gamma(n)=\frac{\displaystyle 1+I (\chi/\chi_{\rm P})}{\displaystyle 1+
\gamma_{\rm imp}\frac{\displaystyle 2E_F}{\displaystyle W_n^2+E_F^2}},
\end{equation}
and the spin susceptibility $\chi$ recovers the usual Stoner-like
expression:
\begin{equation}
\chi = \frac{\chi_0}{1-I (\chi_0/\chi_{\rm P})},
\end{equation}
where the bare spin susceptibility $\chi_0$ is now affected by the
non magnetic impurity scattering:
\begin{equation}
\chi_0 = \chi_{\rm P}
T\sum_{m}
\frac{2E_F}{W_m^2+E_F^2 + \gamma_{\rm imp}2E_F}.
\label{chi0}
\end{equation}
For large Fermi energy systems, $E_{\rm F} \gg \gamma_{\rm imp}$,
equation (\ref{chi0}) reduces to the Pauli spin susceptibility
$\chi_0 = \chi_{\rm P}$. It is thus clear the non magnetic
impurity effects can appear only if the Fermi energy is small enough
to be comparable with $\gamma_{\rm imp}$. Note that in the presence
of electron-phonon interaction an additional energy scale is provided
by $\omega_{\rm ph}$, so that additional anomalous impurity effects are
ruled by the additional parameter $\gamma_{\rm imp} / \omega_{\rm ph}$.

\begin{figure}
\centerline{\psfig{figure=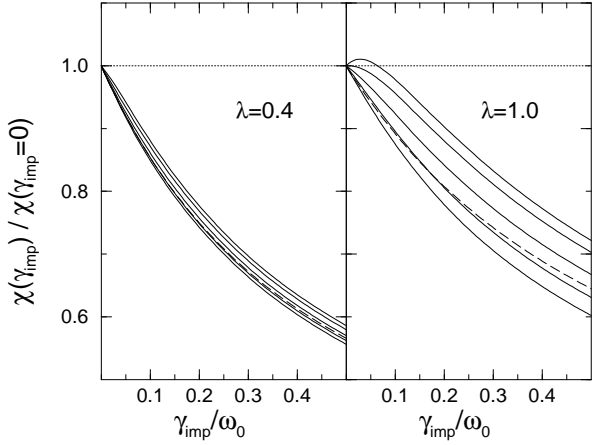,width=8cm}}
\vspace{1mm}
\caption{Spin susceptibility $\chi$ as function of the
impurity scattering rate $\gamma_{\rm imp}$ in the presence of
electron-phonon interaction($\lambda=0.4$, left panel;
$\lambda=1.0$, right panel, $\omega_0/E_{\rm F}=0.7$ and
electron-electron exchange repulsion $I=0.4$.
Solid lines: nonadiabatic vertex corrected theory
with different values of $Q_c$ (from the top to the bottom:
$Q_c=0.1, 0.3, 0.5, 0.7, 0.9$); dashed line: non crossing
approximation.}
\label{f-chi-vs-g}
\end{figure}

In Fig. \ref{f-chi-vs-g} we plot the behaviour of the static
spin susceptibility $\chi$ in a small Fermi energy
system ($\omega_0/E_{\rm F}=0.7$)
as function of the
impurity scattering rate $\gamma_{\rm imp}$.
The data are normalized with respect to the
``pure'' limit $\gamma_{\rm imp} \rightarrow 0$.
Left panel refer to a weak coupling electron-phonon case
($\lambda=0.4$), right panel to strong coupling ($\lambda=1.0$).
In both the case a Stoner factor $I=0.4$ was considered.
Solid lines represent the nonadiabatic vertex corrected theory
with different values of $Q_c$ (from the top to the bottom:
$Q_c=0.1, 0.3, 0.5, 0.7, 0.9$) and the dashed line the non crossing
approximation where only finite bandwidth effects were retained
[$P(Qc;n,m)=C(Q_c;n,m)=0$ in Eqs. (\ref{V1})-(\ref{ave6})].
Fig. \ref{f-chi-vs-g} shows a strong reduction of $\chi$ due to the
impurities scattering with respect to a
large Fermi energy case ($E_{\rm F} \gg \omega_0$, dotted line).
We observe only a weak dependence on the electron-phonon
coupling (right panel data are slightly higher than the left
panel), while the introduction of the nonadiabatic vertex and cross
diagrams essentially leads to a spread of the different curves
for different values of $Q_c$.

\begin{figure}
\centerline{\psfig{figure=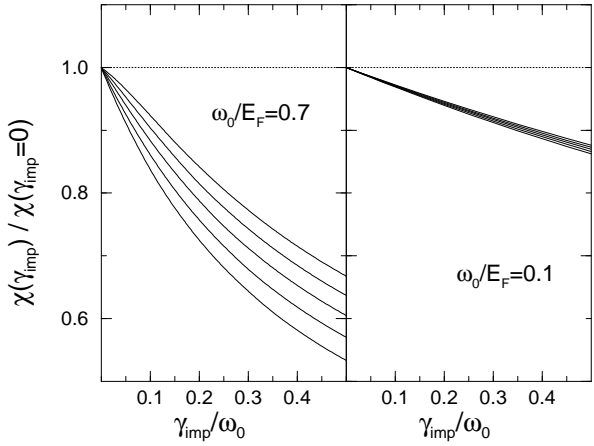,width=8cm}}
\vspace{1mm}
\caption{Spin susceptibility $\chi$ as function of the
impurity scattering rate $\gamma_{\rm imp}$
for a small Fermi energy ($\omega_0/E_{\rm F}=0.7$, left panel)
and for a large Fermi energy
system ($\omega_0/E_{\rm F}=0.1$, right panel).
Solid lines corresponds to different values of $\lambda$:
(from bottom to the top) $\lambda=0.2, 0.4, 0.6, 0.8, 1$.
Other parameters: $Q_c=0.4$ and $I=0.4$.
Dotted line: infinite Fermi energy case.}
\label{f-chi-vs-g2}
\end{figure}

From this behaviour we can argue that the electron-phonon interaction
$\lambda$ plays a secondary role in the reduction of $\chi$ due
to impurity scattering. In similar way a marginal role
is recovered for the electron-electron interaction (Stoner factor $I$).
As matter of facts, the leading effects are ruled by the comparison
between the energy scales $\gamma_{\rm imp}$, $\omega_0$ and
$E_{\rm F}$. In order to highlight this point we
compare in Fig. \ref{f-chi-vs-g2} the dependence of
the magnetic susceptibility $\chi$ on the impurity scattering
rate for a small Fermi energy ($\omega_0/E_{\rm F}=0.7$,left panel)
and for a large Fermi energy
system ($\omega_0/E_{\rm F}=0.1$, right panel).
Solid lines corresponds to different values of $\lambda$
(from bottom to the top): $\lambda=0.2, 0.4, 0.6, 0.8, 1$.
Here we set $Q_c=0.4$ and $I=0.4$ but, as above discussed, different values
would not change the physics.
Note the remarkable difference between left and right panel.
The same impurity scattering rate, which in the presence of a small
Fermi energy
($\omega_0/E_{\rm F}=0.7$) would lead to a reduction of $\chi$
of about 33-47\%, gives rise only to a 12-14\% reduction when
Fermi energy is considerably increased
($\omega_0/E_{\rm F}=0.1$), and to no reduction
at all for the infinite Fermi energy case (dotted line).

Curiously the presence of the electron-phonon interaction
decreases the sensitivity of $\chi$ to impurity scattering rate.
This can be understood considering that the the electron-phonon
scattering reduces by itself the magnetic susceptibility $\chi$ \cite{cgp},
so that further reduction by non magnetic impurity scattering
is disfavored. For the same reasons a stronger Stoner factor
$I$ would enhance the reduction of $\chi$.

\section{Disorder and nonmagnetic impurities in real materials
(MgB$_2$, fullerenes, \ldots)}

We are now in the position to re-address the open questions
arisen in the introduction, concerning namely:
the origin of the
reduction of the density of states as probed by
NMR susceptibility measurements;
the discrepancy between the different behaviour of Mg and B
NMR measurements.
In particular we suggest that the reduction of the 
spin-lattice relaxation rate $1/T_1 T$ upon induced disorder
could reflect the small Fermi energy nature of the electronic
structure probed by the experiments.
Within this context the insensitivity to disorder scattering of
the spin-lattice relaxation rate $1/T_1 T$ probed on $^{25}$Mg in contrast
to the marked reduction of $^{11}$B NMR measurements acquires a natural
explanation related to the different Fermi energy scales involved
in the two cases.
$^{25}$Mg NMR measurements mainly probe the $\pi$ band structures with
high Fermi energy $E_{\rm F} \sim 5$ eV, Using a typical phonon frequency
$\omega_0 \simeq 70$ meV \cite{renker} we estimate
$\omega_0/E_{\rm F}^{\pi} \sim 0.014$ which yields a negligible
dependence of the spin susceptibility on the amount of disorder.
On the other hand NMR on the $^{11}$B boron nucleus is strongly
coupled with the in-plane $\sigma$ orbitals with
Fermi energy $E_{\rm F} \sim 0.4-0.6$ eV. The same phonon
frequency scale gives thus $\omega_0/E_{\rm F}^{\sigma} \sim 0.12-0.14$,
where visible impurity scattering effects are expected.
We conclude that the small Fermi energy of the $\sigma$ bands
is the major responsible for the reduction of
the magnetic susceptibility when probed on $^{11}$B nuclei compared
with NMR measurements on the same quantity on $^{25}$Mg.

Note that the usual two band model within the Migdal-Eliashberg
framework, with different electron-phonon coupling
for $\sigma$ and $\pi$ bands
($\lambda_\sigma \sim 1$, $\lambda_\pi \sim 0.2$),
can not alone explain the discrepancy between the reduction rate
probed by NMR. Indeed: a) electron-phonon interaction
does not affect impurity scattering for high Fermi energy systems
($E_{\rm F} \gg \gamma_{\rm imp}, \omega_0$);
b) the larger electron-phonon coupling constant in the $\sigma$ bands
would predict a smaller reduction rate of $\chi$ as compared
with the smaller $\lambda$ of the $\pi$ bands.

The above discussion suggests that the reduction of $\chi$
upon disorder is a possible tool to point out nonadiabatic effects
in small Fermi energy materials where $\omega_0 \sim E_{\rm F}$.
In this perspective it is interesting to compare the simultaneous
reduction of $\chi$ and $T_c$ as function of the impurity scattering
rate $\gamma_{\rm imp}$. In this way in principle one can trace out
the effects of non magnetic impurity scattering in small Fermi
energy systems as functions of physical measurable quantities
as $\chi$ and $T_c$ avoiding the use of the unaccessible parameter
$\gamma_{\rm imp}$.

\begin{figure}
\centerline{\psfig{figure=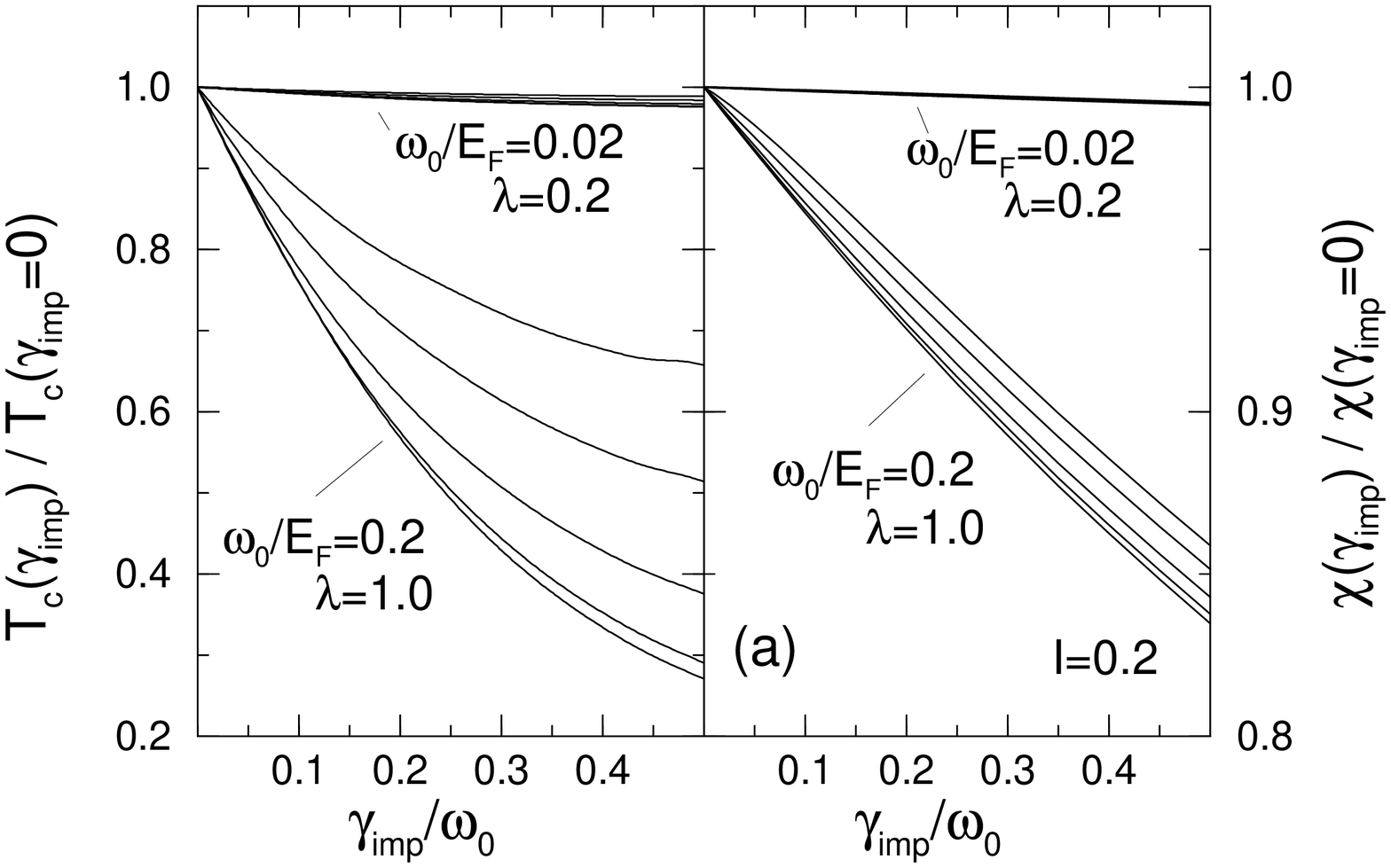,width=8.5cm}}
\vspace{1mm}
\centerline{\psfig{figure=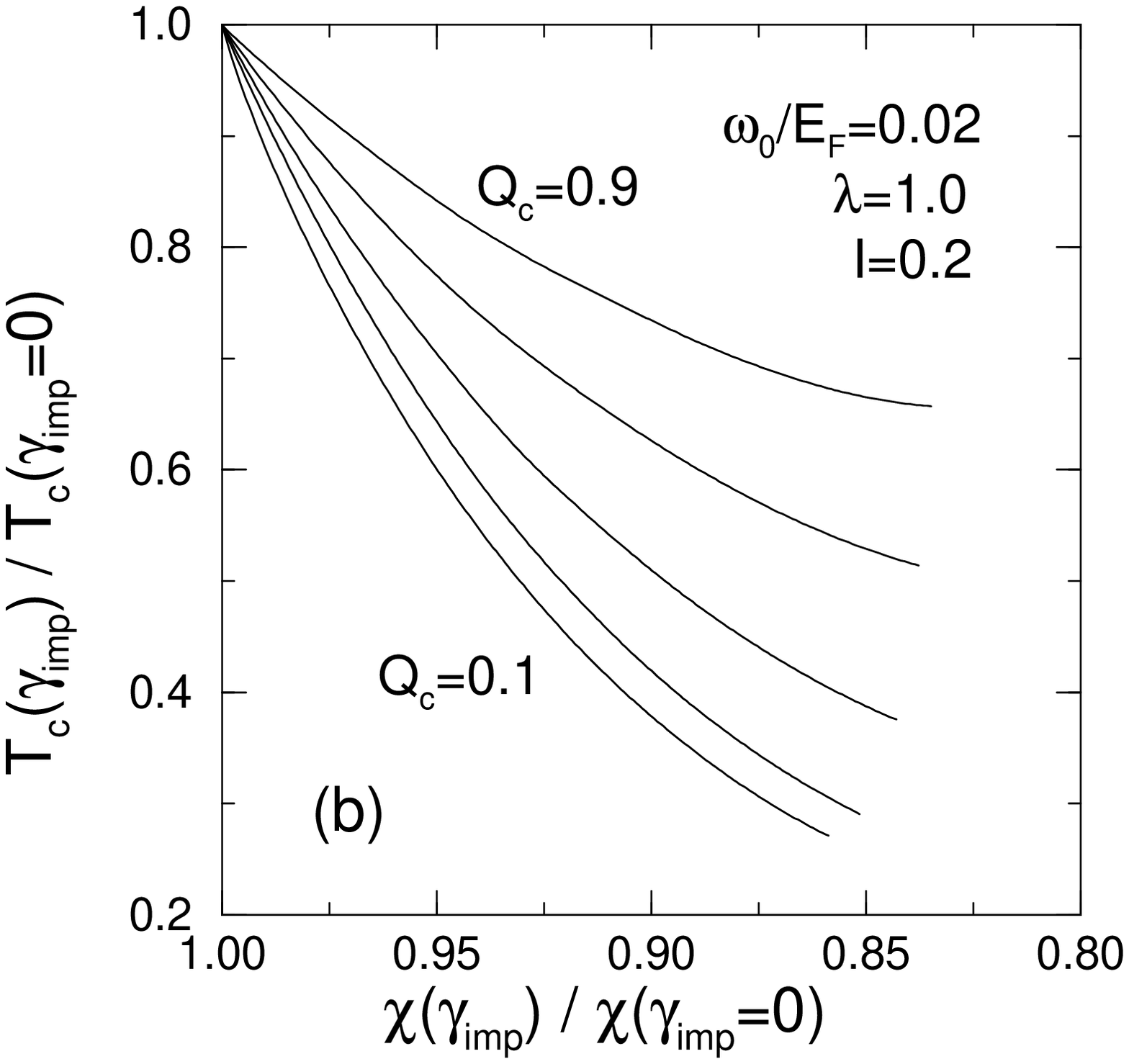,width=6cm}}
\vspace{1mm}
\caption{(a) Reduction of $T_c$ (left panel) and $\chi$
(right panel) as function of $\gamma_{\rm imp}$
for the cases ($\lambda=1.0$, $\omega_0/E_{\rm F}=0.2$, $I=0.2$)
and ($\lambda=0.2$, $\omega_0/E_{\rm F}=0.02$, $I=0.2$),
qualitatively representative respectively of the $\sigma$ and $\pi$ bands
in MgB$_2$. Different lines corresponds to different $Q_c$'s
(from the bottom to the top in left panel, from the top
to the bottom in right panel): $Q_c = 0.1, 0.3, 0.5, 0.7, 0.9$.
(b) Corresponding plot of $T_c$ vs. $\chi$ as varying $\gamma_{\rm imp}$
($\gamma_{\rm imp}=0$ at the left end, 
$\gamma_{\rm imp}=0.5 \omega_0$ at the right end)
for ($\lambda=1.0$, $\omega_0/E_{\rm F}=0.2$, $I=0.2$).
Different values of $Q_c$ are reported as in the previous captions.}
\label{f-tcvschi}
\end{figure}

A conserving derivation of the superconducting equations in a
fully consistent way with the evaluation of the spin susceptibility
follows once again the Baym-Kadanoff theory based on Fig. \ref{f-self}
written in Nambu notation.
A formal derivation of those equations and some technicalities
about the numerical calculations of $T_c$ were discussed in
Ref. \cite{sgp} where we refer for more details.
In Fig. \ref{f-tcvschi}a we plot 
$\chi$ and $T_c$ as functions of $\gamma_{\rm imp}$ for $I=0.2$
and the couples of parameters ($\lambda=1.0$, $\omega_0/E_{\rm F}=0.2$),
($\lambda=0.2$, $\omega_0/E_{\rm F}=0.02$). These cases should
be qualitatively representative of the MgB$_2$ $\sigma$ and $\pi$ bands
which are respectively: strong coupled with small Fermi energy;
and weak coupled with large Fermi energy. Note that a significant
dependence on $\gamma_{\rm imp}$, for both $T_c$ and $\chi$
is observed only for ($\lambda=1.0$, $\omega_0/E_{\rm F}=0.2$)
which represents the case of $\sigma$ bands. The $T_c$ vs. $\chi$ plot
is shown in Fig. \ref{f-tcvschi}b for 
($\lambda=1.0$, $\omega_0/E_{\rm F}=0.2$, $I=0.2$). A reduction of
$T_c$ of the order of $30-80 \%$ is predicted for a reduction
of $\chi$ of $\sim 20 \%$, depending on the parameter $Q_c$.
The general trend is thus in agreement with the experimental
data reported in Ref. \cite{gerashenko}.

In MgB$_2$, where the electronic correlation is thought to be negligible,
there is no reason to expect a significant
momentum selection and $Q_c$ is expected to be $Q_c \sim 1$.
In this situation our analysis would underestimate the suppression of
$T_c$ ($\Delta T_c / T_c \sim 30 \%$) and  $\chi$ 
($\Delta \chi / \chi \sim 20 \%$) when compared with the
experimental scenario, although
some care should be used to extrapolate from the static magnetic
susceptibility $\chi$ to $1/ T T_1$.

It is clear however that
additional ingredients are
required to be taken into account for a quantitative analysis of the
experimental data
of $T_c$ vs. induced disorder.
In particular the discussion in terms of two
separated $\sigma$ and $\pi$ bands is expected to be a poor description 
for the superconducting properties of a complex multiband system as MgB$_2$.
On this basis we conclude that further investigation is needed to
account in a fully satisfactory way for the anomalous dependence
of $T_c$ on the amount disorder. On the other hand
the reduction of the spin susceptibility
upon disorder and non magnetic impurities in MgB$_2$
could be qualitatively understood within the present analysis.
In paricular our results suggest that a primary role
could be played by the small Fermi energy effects driven in MgB$_2$ by the
closeness of the chemical potential to the top
of the $\sigma$ band. In this framework
the different behaviour of B and Mg NMR data receives a natural explanation.

Interesting perspectives are also opened in regards to the fullerene
based materials. The analysis is indeed simplified in these compounds
as a single Fermi energy is present. The extreme smallness of $E_{\rm F}$
is fullerenes ($E_{\rm F} \simeq 0.25$ eV) suggests that disorder
or nonmagnetic impurity effects could lead to even more marked reduction
of $T_c$ and $\chi$ than in MgB$_2$.
A suppression of $T_c$ upon induced disorder as previously been reported
in Ref. \cite{watson}
At our knowledge no measurements of magnetic susceptibility
as function of disorder or impurity amount as been at the present
performed. Experimental work along this line is thus encouraged.

\acknowledgement

This work was partially supported by INFM Research Project
PRA-UMBRA

\endacknowledgement


\begin{thebibliography}{99}
\bibitem{note-localeffects}
We neglect here non local effects appearing in the
neighbouring of a single impurity. In this perspective only
physical quantities averaged on the impurity configuration,
which do not break translational invariance,
will be in the following involved.
\end{thebibliography}
\end{document}